\documentclass [aps,prb,showpacs] {revtex4}
\usepackage[final]{graphics}
\usepackage{amssymb}
\usepackage{amsfonts}
\usepackage{epsfig}

\begin{document}
%\draft
\date{\today}
\title{Quantum Spin-$1$ Anisotropic Ferromagnetic Heisenberg Model in a Crystal Field: A Variational Approach}
\author{D. C. Carvalho$^1$, J. A. Plascak$^{1,2}$ and  L. M. Castro$^3$  }
\affiliation{ $^1$Departamento de F\'\i sica, Instituto de Ci\^encias Exatas,
 Universidade Federal de Minas Gerais, C.P. 702, 
 30123-970 Belo Horizonte, MG - Brazil\\
$^2$ Center for Simulational Physics, University of Georgia, 30602 Athens, Georgia, USA\\
$^3$Departamento de Ci\^encias Exatas, Universidade Estadual do 
Sudoeste da Bahia, Estrada do Bem Querer Km04, CP 95, 45083-900 Vit\'oria da 
Conquista, BA - Brazil}
\begin{abstract}
A variational approach based on Bogoliubov inequality for the free energy
is employed in order to treat the quantum spin-$1$ anisotropic ferromagnetic Heisenberg  
model in the presence of a crystal field. Within the Bogoliubov scheme an
improved pair approximation has been used. The temperature dependent thermodynamic functions
have been obtained and provide much better results than the previous simple mean-field
scheme. 
%in which the latter procedure does not distinguish the dimensionality of the %lattice. 
In one dimension, which is still non-integrable for 
quantum spin-$1$,
we get the exact results in the classical limit, or near-exact results in the
quantum case, for the free energy, magnetization and
quadrupole moment, as well for the transition temperature. In two and three dimensions
the corresponding global phase diagrams have been obtained as a function 
of the parameters of the Hamiltonian. First-order transition lines, 
second-order transition lines, tricritical and tetracritical points, and critical
endpoints  have been located
through the analysis of the minimum of the Helmholtz free energy and a Landau like expansion 
in the approximated free energy. 
Only first-order quantum transitions have been found at zero temperature.
Limiting cases, such as isotropic Heisenberg, Blume-Capel and Ising models      
have been analyzed and compared to previous results obtained from other analytical 
approaches as well as from Monte Carlo simulations.
\end{abstract} 
\maketitle
\section{ Introduction}
\label{intro}
Quantum phase transitions have been extensively studied in the literature \cite{ncomm,PRB-85,PRL-93, Hertz, Sond, Kaden, Carr} and their fully 
understanding is still one of the most interesting and important subjects in the modern condensed matter physics, both experimentally and theoretically.
These transitions have been observed in several experimental realizations such as, for instance, the magnetic insulators $LiHoF_{4}$\cite{Bitko}, 
$La_2CuO_4$\cite{Kei}, and heavy-fermion systems\cite{JPS} as $CeRu_2Si_2$, $\beta$-$YbAlB_4$. These quantum phase transitions are driven 
by quantum fluctuations coming from the Heisenberg uncertainty principle (usually due to the existing competition of a field with the ordering energy 
interaction), instead of the classical phase transitions that are just driven by thermal fluctuations (temperature). Despite the quantum phase 
transitions occur only at zero temperature, and then, in a region of difficult experimental access, its effects can also be seen in a finite 
temperature region \cite{sachdev}. Hence, it is very important to study, besides the quantum phase transitions themselves at absolute zero temperature, 
the corresponding phase transitions in the region of low temperatures, in which the quantum effects are certainly still present.

From the theoretical point of view, the simplest non-trivial magnetic model that exhibits quantum phase transitions is the Ising model with a 
transverse field or, simply, quantum transverse Ising model. It is the transverse field that competes with the ordering exchange interaction energy. In this case, 
only the spin-$1/2$ one-dimensional version\cite{pfeuty,Shira} can be exactly solved. In addition, this one-dimensional quantum model can be mapped onto a 
two-dimensional 
classical Ising model. In general, one has indeed that any $d$-dimensional quantum system can be mapped onto an analogous $(d+1)$-classical model\cite{kogut}.

Another important system, and much richer than the quantum transverse Ising model, is the isotropic Heisenberg model. It has been studied for many years, 
both in its classical\cite{Diego,tonico} and quantum versions\cite{tonico-spin1,condens}. Nevertheless, this model can be exactly solved only in its 
classical one-dimensional version\cite{fisher} and in its quantum spin-$1/2$ one-dimensional version\cite{bethe} as well. On the other hand, 
this model, for spin-$1$, in one dimension, 
is not integrable due to the difficulty involved with tackling the non-commutativity of spin operators. Moreover, due a theorem by Mermin and Wagner 
\cite{Mermin}, it has been shown that it is not possible for this system, in one and two dimensions, to present any long-range order at finite 
temperature.

It is known, however, that in realistic systems one expects to find some degree of anisotropy which can modify the global symmetry of the material, 
creating thus axis, or even planes, of easy magnetization \cite{v1,v2,hb,crow}. For instance, the ferromagnetic superconductor  $UGe_2$ \cite{PRL-89} 
exhibits an easy axis anisotropy. Therefore, in any theoretical model, one must consider such features in the Hamiltonian that should describe the 
phenomenon under study. A suitable model that takes into account such asymmetries is the so called {\it anisotropic} Heisenberg model in the 
presence of a crystal field (or single ion anisotropy), which can be written as
\begin{equation}
\label{ham}
{\cal H} = -J\sum_{<i,j>}\left[\eta\left(S_{i}^{x}S_{j}^{x}+S_{i}^{y}S_{j}^{y}\right) + S_{i}^{z}S_{j}^{z}\right]
-D\sum_{i=1}^N{S_{i}^{z}}^{2} - H\sum_{i=1}^NS_{i}^{z},
\end{equation}
where $J>0$ is the ferromagnetic exchange interaction between spins $i$ and $j$, the first sum is over pairs of nearest neighbor sites $<i,j>$, 
and $N$ is the total number of sites of the lattice. The parameter $\eta$  measures the degree of the spin interaction anisotropy (this model is also called XXZ model) in the region 
of easy-axis for $\eta<1$, or easy-plane for $\eta>1$ , and $D$ plays the role of the crystal field. $H$ is 
the external magnetic field which will be set to zero and $S_{i}^{\alpha}$ are the  $\alpha=x,y,z$ components of spin operators at site $i$ with 
the eigenvalues of $S_{i}^{z}$ operator taking the values $-S, -S+1,..., S-1, S$. 

The above model has some interesting limits, namely: ({\it i}) for $\eta=0$ it reduces to the classical Blume-Capel model with general spin-$S$; 
({\it ii}) for $\eta=1$, $D=0$ and $H=0$ one has the isotropic Heisenberg model; and ({\it iii}) when $D\rightarrow \infty$ the Hamiltonian is 
equivalent to the spin-$1/2$ Ising model. The classical model in item ({\it i}) has been extensively studied, for instance, by mean-field 
approximations \cite{b,c,pla,lara1}, mean-field renormalization group \cite{lara2}, Monte Carlo simulations \cite{jl,pd,capa}, conformal 
invariance \cite{xapp}, among others. The phase diagram consists of ordered and disordered phases separated by second- and first-order transition lines, with tricritical and double critical end 
points for integer values of $S$ (except for $S=1$ which has only one tricritical
point), and only double critical end points for semi-integer values of the spin $S$.

We will consider herein spin $S=1$ in order to study the crystal field effects  on the quantum model when $\eta\ne0$. In particular, it will also be 
interesting to better understand how the quantum fluctuations will affect the presence of the tricritical points, mainly for the three-dimensional 
lattice. It should be stressed that experimental realizations of spin-$1$ systems range from metamagnet \cite{schmidt}, magnetic materials 
(see, for instance, reference \cite{oitmaa} and references therein) to He$^3$-He$^4$ mixtures \cite{beg}. On the other hand, from the 
theoretical point of view, in the $\eta=1$ case, limit ({\it ii}) above, the spin-$1$ ferromagnetic isotropic Heisenberg model in the 
presence of an arbitrary crystal-field potential has been treated by mean-field approximation \cite{mfa,khaje} and a linked-cluster expansion 
method \cite{Kok-PRB}. However, to the best of our knowledge, the complete Hamiltonian (\ref{ham}) with spin one has been treated only 
by a mean-field like approach, which does not distinguish neither the topology nor the dimension of the lattice \cite{khaje,chinese}. Moreover, 
the topologies of the corresponding phase diagrams have not been detailed enough to give a clear picture of all the transitions involved, 
mainly the quantum phase transitions at zero temperature.  It would be worthwhile thus to investigate the behavior of this anisotropic 
Heisenberg model with a crystal field by taking a better, or more reliable, approach, even in its one-dimensional version. 
The procedure we will follow is closely related to 
the variational approach based on Bogoliubov inequality for the free energy \cite{falk}, within the pair approximation \cite{Ferreira}, 
which reproduces exact results in some limiting cases.

The plan of the paper is as follows. In the next section, we present the theoretical approach for getting the free energy and the thermodynamic quantities of interest. In
section \ref{res} we present the numerical results. Some concluding remarks are given
in the last section, and in the Appendix some of the analytical equations are presented.
\section{Variational Approach for the free energy}
\label{ham-var}
The pair variational procedure and the corresponding thermodynamic functions of interest will be presented below. The potentiality of the 
approximation will be discussed by comparing the results in some limiting cases, where exact or more reliable approaches have been 
previously employed.

\subsection{Bogoliubov Variational Approach}

The variational approach that will be employed is based on Bogoliubov inequality for the free energy \cite{falk}
\begin{equation} \label{bi}
F\leq F_{0}+\langle {\cal H}-{\cal H}_0(\gamma)
\rangle_{0}\equiv \Phi(\gamma) ,
\end{equation}
where ${\cal H}$ is the Hamiltonian under study (\ref{ham}),
${\cal H}_0(\gamma)$ is a trial Hamiltonian which can be 
exactly solved and depends on variational parameters designated by $\gamma$. $F$ is the free 
energy of the system described by $\cal H$, $F_0$ is the corresponding free energy of the
trial Hamiltonian ${\cal H}_0$, and the thermal average $<... >_0$ is taken
over the ensemble defined by ${\cal H}_0$. The
approximate free energy is then given by the minimum of $\Phi(\gamma)$ with
respect to $\gamma$, i.e. $F\equiv \Phi_{min}(\gamma)$.

We will follow herein the pair approximation by Ferreira et al\cite{Ferreira} consisting of taking $n_1$ single free spins and $n_2$ disconnected pairs of spins 
distributed on the lattice, in such a way that $n_1+2n_2=N$. 
As the Hamiltonian (\ref{ham}) has, in principle, either easy axis (for $\eta<1$) or 
easy plane (for $\eta>1$) ordering, each term of the above trial 
Hamiltonian can be chosen as a sum of two parts
%
%\begin{equation}
% \label{trial-gen}
%{\cal H}_{0}={\cal H}_{0\parallel}+ {\cal H}_{0\perp},
%\end{equation}
%
%Each of the terms in eq. (\ref{trial-gen}), in their turn, can be written as
%
\begin{equation}
 \label{trial-free}
{\cal H}_{0}^f={\cal H}_{0\parallel}^{f}+{\cal H}_{0\perp}^{f},
\end{equation}
\begin{equation}
 \label{trial-pair}
{\cal H}_{0}^p={\cal H}_{0\parallel}^{p}+{\cal H}_{0\perp}^{p},
\end{equation}
where ${\cal H}_{0\parallel}^{f}$ and ${\cal H}_{0\parallel}^{p}$ take into account, respectively, the free and pairs of spins ordering along the $z$ axis, 
and ${\cal H}_{0\perp}^{f}$ and ${\cal H}_{0\perp}^{p}$ take into account the corresponding ordering along the $xy$ plane. 
The free and pairs of spins Hamiltonian components can be then written as 
\begin{equation}
 \label{equa-hfp}
{\cal H}_{0\parallel}^{f}=-{\sum_{free}}\left[h_{1}^{\parallel}S_{i}^{z}+D{S_{i}^{z}}^{2}+{\gamma}_{1}^{\parallel}{S_{i}^{z}}^{2}\right],
\end{equation}
\begin{equation}
 \label{equa-hpp}
{\cal H}_{0\parallel}^{p}=-\sum_{pair}\left\{ J\left[\eta{\vec S_{i}}\cdot{\vec S_{j}}+(1-\eta)S_{i}^{z}S_{j}^{z}\right] + 
D\left({{S_{i}^{z}}}^{2}+{{S_{j}^{z}}}^{2}\right)
+h_{2}^{\parallel}\left({S_{i}^{z}}+{S_{j}^{z}}\right)
+\gamma_{2}^{\parallel}\left({{S_{i}^{z}}}^{2}+{{S_{j}^{z}}}^{2}\right)\right\},
\end{equation}
\begin{equation}
 \label{equa-hperf}
{\cal H}_{0\bot}^{f}=-{\sum_{free}}\left\{\frac{h_{1}^{\bot}}{\sqrt{2}}(S_{i}^{x}+S_{i}^{y})+D{S_{i}^{z}}^{2}+
{\gamma}_{1}^{\bot}{S_{i}^{z}}^{2}\right\},
\end{equation}
\begin{equation}
\label{equa-hperp}
{\cal H}_{0\bot}^{p}=-\sum_{pair}\left\{J\left[\eta{\vec S_{i}}\cdot{\vec S_{j}}+(1-\eta)S_{i}^{z}S_{j}^{z}\right] + 
D\left({{S_{i}^{z}}}^{2}+{{S_{j}^{z}}}^{2}\right)
+\frac{h_{2}^{\bot}}{\sqrt{2}}\left({S_{i}^{x}}+{S_{j}^{x}}+{S_{i}^{y}}+{S_{j}^{y}}\right)
+\gamma_{2}^{\bot}\left({{S_{i}^{z}}}^{2}+{{S_{j}^{z}}}^{2}\right)\right\},
\end{equation}
where $h_{1}^{\parallel}$, $h_{2}^{\parallel}$, $\gamma_1^{\parallel}$ and $\gamma_2^{\parallel}$ are variational parameters 
along the parallel direction of the $z$ axis and $h_{1}^{\bot}$, $h_{2}^{\bot}$, $\gamma_1^{\bot}$ and $\gamma_2^{\bot}$ are 
variational parameters in the $xy$ plane. The sum $\sum_{free}$ is taken over all isolated spins and $\sum_{pair}$ is taken 
over all disconnected pairs of spins.
A similar choice for the trial Hamiltonian has been proposed by Lara {\it et al}\cite{lara1} in the study of the classical 
Blume-Capel model, 
corresponding to the limiting case $\eta=0$ in our Hamiltonian (\ref{ham}). In the present paper, we have generalized such 
trial Hamiltonian for different values of the anisotropy $\eta$, therefore, allowing for the presence of quantum fluctuations 
in the system, which significantly complicates the analysis.  In this pair approximation two nearest-neighbor spins fluctuations
are taken into account exactly, while in the previous usual mean-field approach no fluctuations at all have been considered\cite{khaje}.

From the trial Hamiltonian ${\cal H}_{0}$, we can write the partition function ${\cal Z}_{0}$ as
\begin{equation}
{\cal Z}_0=Tre^{-\beta{\cal H}_0}=Tre^{
-\beta({\cal H}_{0}^f+ {\cal H}_{0}^p)}=Tre^{-\beta({\cal H}_{0\parallel}^{f}+{\cal H}_{0\parallel}^{p}+
{\cal H}_{0\bot}^{f}+{\cal H}_{0\bot}^{p})}={\cal Z}_{0\parallel}^{f}{\cal Z}_{0\parallel}^{p}{\cal Z}_{0\bot}^{f}{\cal Z}_{0\bot}^{p},
\end{equation}
in which the free Hamiltonian contributions for the partition function are given by
\begin{eqnarray}
{ \cal Z}_{0\parallel}^{f}&=&Tre^{-\beta{\cal H}_{0\parallel}^{f}} = \left({\cal Z}_{1\parallel}^{f}\right)^{n_1},\nonumber\\
{ \cal Z}_{0\bot}^{f}&=&Tre^{-\beta{\cal H}_{0\bot}^{f}} = \left({\cal Z}_{1\bot}^{f}\right)^{n_1},\nonumber
\end{eqnarray}
where $\beta=1/k_BT$, with $k_B$ the Boltzmann constant. 
The one-spin $3\times3$ Hamiltonian matrix can be easily diagonalized yielding
\begin{eqnarray}
{\cal Z}_{1\parallel}^{f}&=&1+2e^{\beta(D+{\gamma}_1^{\parallel})}\cosh(\beta{h_1^{\parallel}}), \label{z1fpl}\\
{\cal Z}_{1\bot}^{f}&=&e^{\beta(D+{\gamma}_1^{\bot})} + 2e^{\frac{\beta(D+{\gamma}_1^{\bot})}{2}}\cosh\left(\frac{\beta\sqrt{(D+\gamma_1^{\bot})^2 + 
4(h_{1}^{\bot})^2}}{2}\right).
\label{z1fpp}
\end{eqnarray}

Analogously, for the parallel component of the pair Hamiltonian we get
\begin{equation}
 {\cal Z}_{0\parallel}^{p}=Tre^{-\beta{\cal H}_{0\parallel}^{p}} = \left({\cal Z}_{2\parallel}^{p}\right)^{n_2},
\end{equation}
\begin{equation}
 {\cal Z}_{2\parallel}^{p}=4e^{\beta(D+\gamma_2^{\parallel})}\cosh(\beta{h_2^{\parallel}})\cosh(\beta{J\eta})+e^{-\beta[J-2(D+\gamma_2^{\parallel})]}
+2e^{\beta[J+2(D+{\gamma}_2^{\parallel})]}\cosh(2\beta{h_2^{\parallel}}) + 2e^{-\frac{\beta\alpha}{2}}\cosh\left(\frac{\beta\Delta}{2}\right),
\end{equation}
\begin{eqnarray}
\alpha=J-2(D+\gamma_2^{\parallel})~~~\mbox{and}~~~ \Delta=\sqrt{\alpha^2+8(J\eta)^2}.\nonumber
\end{eqnarray}
The above expression comes from the parallel pair Hamiltonian which is a $9\times9$ matrix that can be analytically diagonalized. 
A different situation, however, holds for the perpendicular ($xy$ plane) component of the pair Hamiltonian. In this case we can not
obtain an analytical expression for the nine eigenvalues of the corresponding Hamiltonian and we have to resort to a numerical 
diagonalization of  ${\cal H}_{0\bot}^{p}$ in order to get the partition function ${ \cal Z}_{0\bot}^{p}$.

After calculating the terms appearing in the Bogoliubov inequality, we can write the free energy per particle as 
\begin{eqnarray}
 f=\frac{F}{N} = f_\parallel + f_\bot =\frac{F_{\parallel}}{N}+\frac{F_{\bot}}{N},
\label{fe}
\end{eqnarray}
where the free energies $f_{\parallel}$ and $f_{\bot}$ are given by
\begin{eqnarray}
\label{energy-para}
f_{\parallel}=\frac{F_{\parallel}}{N}&=&-\frac{c}{2}k_{B}T\ln{{\cal Z}_{2\parallel}^{p}}+(c-1)k_{B}T\ln{{\cal Z}_{1\parallel}^{f}}
+(1-c)(h_{1}^{\parallel}m_{\parallel}+\gamma_{1}^{\parallel}q_{\parallel})+c(h_{2}^{\parallel}m_{\parallel}+\gamma_{2}^{\parallel}q_{\parallel}),
\end{eqnarray}
\begin{eqnarray}
\label{energy-perp}
f_{\bot}=\frac{F_{\bot}}{N}&=&-\frac{c}{2}k_{B}T\ln{{\cal Z}_{2\bot}^{p}}+(c-1)k_{B}T\ln{{\cal Z}_{1\bot}^{f}}
+(1-c)(h_{1}^{\bot}m_{\bot}+\gamma_{1}^{\bot}q_{\bot})+c(h_{2}^{\bot}m_{\bot}+\gamma_{2}^{\bot}q_{\bot}),
\end{eqnarray}
where $c$ is the coordination number of the lattice.
In the above equations $m_{\parallel}$ and $m_{\bot}$ are the parallel and perpendicular components of the magnetization
defined by
\begin{equation}
 \label{mpl}
m_{\parallel}\equiv\langle{S_{i}^{z}}{\rangle}_{0}=\frac{1}{\beta}\frac{\partial{\ln{\cal Z}_{1\parallel}^{f}}}{\partial{h_{1}^{\parallel}}}=
\frac{1}{2\beta}\frac{\partial{\ln{\cal Z}_{2\parallel}^{p}}}{\partial{h_{2}^{\parallel}}},
\end{equation}
\begin{equation}
 \label{mpp}
m_{\bot}\equiv\sqrt{\langle{S_{i}^{x}}{\rangle}_{0}^{2} +\langle{S_{i}^{y}}{\rangle}_{0}^{2}}=\frac{1}{\beta}
\frac{\partial{\ln{\cal Z}_{1\bot}^{f}}}{\partial{h_{1}^{\bot}}}=
\frac{1}{2\beta}\frac{\partial{\ln{\cal Z}_{2\bot}^{p}}}{\partial{h_{2}^{\bot}}}.
\end{equation}
Note from Eqs. (\ref{mpl}) and (\ref{mpp}) that the magnetization coming from single spins and from pairs of spins are
the same in order to keep the translational symmetry of the model. Similarly, we get for the parallel and perpendicular
components of the quadrupole moments $q_{\parallel}$ and $q_{\bot}$
\begin{equation}
\label{qpl}
q_{\parallel}\equiv\langle{S_{i}^{z}}^{2}\rangle_{0}=\frac{1}{\beta}\frac{\partial{\ln{\cal Z}_{1\parallel}^{f}}}{\partial{\gamma_{1}^{\parallel}}}=
\frac{1}{2\beta}\frac{\partial{\ln{\cal Z}_{2\parallel}^{p}}}
{\partial{\gamma_{2}^{\parallel}}},
\end{equation}
\begin{equation}
\label{qpp}
q_{\bot}\equiv\langle{S_{i}^{z}}^{2}\rangle_{0}=\frac{1}{\beta}\frac{\partial{\ln{\cal Z}_{1\bot}^{f}}}{\partial{\gamma_{1}^{\bot}}}
=\frac{1}{2\beta}\frac{\partial{\ln{\cal Z}_{2\bot}^{p}}}
{\partial{\gamma_{2}^{\bot}}}.
\end{equation}

After minimizing the free-energy $f$ with respect to the eight variational parameters $h_{1}^{\parallel}$, $h_{2}^{\parallel}$, 
$\gamma_1^{\parallel}$ and $\gamma_2^{\parallel}$, and $h_{1}^{\bot}$, $h_{2}^{\bot}$, $\gamma_1^{\bot}$ and $\gamma_2^{\bot}$, 
we obtain the following relations 
\begin{eqnarray}
\label{h1-h2}
(c-1) h_1^{\parallel}=ch_2^{\parallel}~~~\mbox{and}~~~(c-1)\gamma_1^{\parallel}=c\gamma_2^{\parallel},
\end{eqnarray}
\begin{eqnarray}
\label{h1-h2-perp}
(c-1) h_1^{\bot}=ch_2^{\bot}~~~\mbox{and}~~~(c-1)\gamma_1^{\bot}=c\gamma_2^{\bot}.
\end{eqnarray}
Thus, for a given value of the Hamiltonian parameters $D/J$, $\eta$ and $c$, and at a reduced temperature $t=\frac{k_{B}T}{J}$, one
can solve Eqs. (\ref{mpl})-(\ref{h1-h2-perp}) in order to get the dimensionless reduced variational parameters 
 $h_{1}^{\parallel}/J$, $h_{2}^{\parallel}/J$, 
$\gamma_1^{\parallel}/J$ and $\gamma_2^{\parallel}/J$, and $h_{1}^{\bot}/J$, $h_{2}^{\bot}/J$, $\gamma_1^{\bot}/J$ and $\gamma_2^{\bot}/J$.
When more than one set of solutions are found, the stable solutions will be those which minimize the approximated free energy. From this procedure
all thermodynamic properties of the system can be computed.

It turns out that the system is only ordered either along the $z$ direction or in the $xy$ plane, in such a way that the variational 
parameters along $z$ and perpendicular to $z$ are decoupled. This allows one to get some analytical results for the critical lines,
tricritical and tetracritical points. For instance, when the perpendicular variational parameters vanish, Eqs. (\ref{mpl}) 
and (\ref{qpl}) yield
\begin{eqnarray}
 \label{mag-para}
\frac{e^{\beta\gamma_{1}^{\parallel}}\sinh(\beta{h_{1}^{\parallel}})}{{\cal Z}_{1\parallel}^{f}}&=&
\frac{e^{\beta\gamma_{2}^{\parallel}}\sinh(\beta{h_{2}^{\parallel}})}{{\cal Z}_{2\parallel}^{p}}
[2e^{\beta(J+D+\gamma_{2}^{\parallel})}\cosh(\beta{h_{2}^{\parallel}}) 
+\cosh(\beta{J}\eta)],
\end{eqnarray}
\begin{eqnarray}
\label{quad-para}
\frac{2e^{\beta(D+\gamma_{1}^{\parallel})}\cosh(\beta{h}_{1}^{\parallel})}{{\cal Z}_{1\parallel}^{f}}&=&
\frac{e^{\beta(D+\gamma_{2}^{\parallel})}}{{\cal Z}_{2\parallel}^{p}}\Big\{2\cosh(\beta{h}_{2}^{\parallel})\cosh(\beta{J}\eta)
+ e^{\beta(J+D+\gamma_{2}^{\parallel})}\left[2\cosh(2\beta{h}_{2}^{\parallel})+e^{-2\beta{J}}\right]\nonumber\\
&+&e^{-\frac{\beta{J}}{2}}\left[\cosh\left(\frac{\beta\Delta}{2}\right)-\frac{\alpha}{\Delta}\sinh\left(\frac{\beta\Delta}{2}\right)\right]\bigg\},
\end{eqnarray}
which together with the equations (\ref{h1-h2}) can be numerically resolved for $h_1^{\parallel}(t)/J$, $h_2^{\parallel}(t)/J$, 
$\gamma_1^{\parallel}(t)/J$ and $\gamma_2^{\parallel}(t)/J$, as a function of the reduced temperature $t$ for a given set of 
Hamiltonian parameters. This gives the ordering of the parallel order-parameter $m_\parallel$ and the thermodynamics of
the parallel ordered phase.

At criticality, equations (\ref{mag-para}) and (\ref{quad-para}) can be simplified, because the magnetization along the $z$ axis 
continuously goes to zero, i.e. $m_\parallel\rightarrow0$,  which is equivalent to take the 
limit $h_1^{\parallel}\rightarrow0$ and $h_2^{\parallel}\rightarrow0$. Hence, we arrive at the following coupled equations for the critical 
temperature of the parallel order parameter 
\begin{eqnarray}
\frac{e^{\beta\gamma_{1}^{\parallel}}z}{(z-1){\cal Z}_{1\parallel}^{f}(0)}&=&\frac{e^{\beta\gamma_{2}^{\parallel}}}{{\cal Z}_{2\parallel}^{p}(0)}
\left[2e^{\beta(J+D+\gamma_{2}^{\parallel})}
+\cosh(\beta{J}\eta)\right] \nonumber 
\end{eqnarray}
and
\begin{eqnarray}
\frac{2e^{\beta(D+\gamma_{1}^{\parallel})}}{{\cal Z}_{1\parallel}^{f}(0)}&=&\frac{e^{\beta(D+\gamma_{2}^{\parallel})}}{{\cal Z}_{2\parallel}^{p}(0)}
\bigg\{2\cosh(\beta{J}\eta)
+ e^{\beta(J+D+\gamma_{2}^{\parallel})}[2+e^{-2\beta{J}}]
+e^{-\frac{\beta{J}}{2}}\left[\cosh\left(\frac{\beta\Delta}{2}\right)-\frac{\alpha}{\Delta}\sinh\left(\frac{\beta\Delta}{2}\right)\right]\bigg\},\nonumber
\end{eqnarray}
where 
\begin{equation}
\label{critic-partition-para-f}
 {\cal Z}_{1\parallel}^{f}(0)=1+2e^{\beta(D+\gamma_1^{\parallel})},
\end{equation}
\begin{eqnarray}
\label{critic-partition-para-p}
 {\cal Z}_{2\parallel}^{p}(0)&=&4e^{\beta(D+\gamma_2^{\parallel})}\cosh(\beta{J\eta})+e^{-\beta[J-2(D+\gamma_2^{\parallel})]}
+2e^{\beta[J+2(D+{\gamma}_2^{\parallel})]} + 2e^{-\frac{\beta\alpha}{2}}\cosh\left(\frac{\beta\Delta}{2}\right).
\end{eqnarray}

Analogously, for the perpendicular plane the same method can be realized to get the perpendicular variational parameters,
since in this case the parallel ones vanish. 
The expressions for $\displaystyle\frac{1}{\beta}\frac{\partial{\ln{\cal Z}_{1\bot}^{f}}}{\partial{h_{1}^{\bot}}}$ and 
$\displaystyle\frac{1}{\beta}\frac{\partial{\ln{\cal Z}_{1\bot}^{f}}}{\partial{\gamma_{1}^{\bot}}}$ can be readily
obtained from Eq. (\ref{z1fpp}) as follows
\begin{eqnarray}
 \label{mag-f-per}
\frac{1}{\beta}\frac{\partial{\ln{\cal Z}_{1\bot}^{f}}}{\partial{h_{1}^{\bot}}}&=&
\frac{4e^{\frac{\beta(D+\gamma_{1}^{\bot})}{2}}\sinh\left(\frac{\beta\sqrt{(D+\gamma_1^{\bot})^2 + 4(h_{1}^{\bot})^2}}{2}\right)\frac{h_{1}^{\bot}}
{\sqrt{(D+\gamma_1^{\bot})^2 + 4(h_{1}^{\bot})^2}}}{{\cal Z}_{1\bot}^{f}},
\end{eqnarray}
\begin{eqnarray}
\label{quad-f-per}
\frac{1}{\beta}\frac{\partial{\ln{\cal Z}_{1\bot}^{f}}}{\partial{\gamma_{1}^{\bot}}}&=&
\frac{\left[e^{\beta(D+\gamma_1^{\bot})}+e^{\frac{\beta(D+\gamma_1^{\bot})}{2}}\left(\cosh(\frac{\beta\sqrt{(D+\gamma_1^{\bot})^2 + 4(h_{1}^{\bot})^2}}{2})
+\frac{(D+\gamma_1^{\bot})\sinh(\frac{\beta\sqrt{(D+\gamma_1^{\bot})^2 + 4(h_{1}^{\bot})^2}}{2})}
{\sqrt{(D+\gamma_1^{\bot})^2 + 4(h_{1}^{\bot})^2}}\right)\right]}{{\cal Z}_{1\bot}^{f}}.
\end{eqnarray}

Nevertheless, as previously stressed, the pair perpendicular Hamiltonian, ${\cal H}_{0\bot}^{p}$, could not be solved analytically, 
meaning we do not have any
analytical expression for $\displaystyle\frac{1}{2\beta}\frac{\partial{\ln{\cal Z}_{2\bot}^{p}}}{\partial{h_{2}^{\bot}}}$ and 
$\displaystyle\frac{1}{2\beta}\frac{\partial{\ln{\cal Z}_{2\bot}^{p}}}{\partial{\gamma_{2}^{\bot}}}$. Everything must be done 
numerically for finite values of $h_1^{\bot}$ as $h_2^{\bot}$. However, at criticality, $h_1^{\bot}$ and $h_2^{\bot}$ go to zero, 
which permit simplifying expressions (\ref{mag-f-per}) and (\ref{quad-f-per}), 
as well as the pair Hamiltonian can be analytically
diagonalized for $h_2^{\bot}=0$. So, by using the usual time independent quantum mechanics perturbation theory up to second 
order in $h_2^{\bot}$, we can get the corresponding expanded eigenvalues. In this way we can write the following coupled equations
\begin{equation}
\label{Tchpp1}
\frac{2c\left[e^{\beta(D+\gamma_{1}^{\bot})}-1\right]}{(1-c)\left(D+\gamma_1^{\bot}\right){\cal Z}_{1\bot}^{f}(0)}=
\frac{\sum_{i}e^{-\beta\delta_{i}}\Gamma_{i}}{{\cal Z}_{2\bot}^{p}(0)},
\end{equation}
\begin{equation}
\frac{2e^{\beta(D+\gamma_{1}^{\bot})}}{{\cal Z}_{1\bot}^{f}(0)}=\frac{e^{\beta(D+\gamma_{2}^{\bot})}}{{\cal Z}_{2\bot}^{p}(0)}
\left\{2\cosh(\beta{J}\eta)
+ e^{\beta(J+D+\gamma_{2}^{\bot})}\left[2+e^{-2\beta{J}}\right]
+e^{-\frac{\beta{J}}{2}}\left[\cosh\left(\frac{\beta\Pi}{2}\right)-\frac{\kappa}{\Pi}\sinh\left(\frac{\beta\Pi}{2}\right)\right]\right\},
\label{Tchpp2}
\end{equation}
where 
\begin{eqnarray}
\label{def-1}
 \sum_{i}e^{-\beta\delta_{i}}\Gamma_{i}&=& \frac{e^{-\beta[J-2(D+\gamma_2^{\bot})]}}{J(1-\eta)-(D+\gamma_2^{\bot})}
+2\frac{e^{\beta[J+2(D+\gamma_2^{\bot})]}}{-J(1-\eta)-(D+\gamma_2^{\bot})}
+\frac{e^{-\beta(J\eta-(D+\gamma_2^{\bot}))}}{-J(1-\eta)+(D+\gamma_2^{\bot})}\nonumber\\
&+&3\frac{e^{-\beta\lambda_{+}}x_{+}^2}{\lambda_{+}+J\eta+D+\gamma_2^{\bot}}
+3\frac{e^{-\beta\lambda_{-}}x_{-}^2}{\lambda_{-}+J\eta+D+\gamma_2^{\bot}}
+2\frac{e^{\beta(J\eta+D+\gamma_2^{\bot})}}{J(1-\eta)+D+\gamma_2^{\bot}}\nonumber\\
&+&3x_{+}^2\frac{e^{\beta(J\eta+D+\gamma_2^{\bot})}}{-(J\eta+D+\gamma_2^{\bot})-\lambda_{+}}
+3x_{-}^2\frac{e^{\beta(J\eta+D+\gamma_2^{\bot})}}{-(J\eta+D+\gamma_2^{\bot})-\lambda_{-}}, 
\end{eqnarray}
\begin{eqnarray}
\label{critic-partition-perp-f}
 {\cal Z}_{1\bot}^{f}(0)&=&1+2e^{\beta(D+\gamma_1^{\bot})},
\end{eqnarray}
\begin{eqnarray}
\label{critic-partition-perp-p}
{\cal Z}_{2\bot}^{p}(0)&=&4e^{\beta(D+\gamma_2^{\bot})}\cosh(\beta{J\eta})+e^{-\beta[J-2(D+\gamma_2^{\bot})]}
+2e^{\beta[J+2(D+{\gamma}_2^{\bot})]} + 2e^{-\frac{\beta\kappa}{2}}\cosh\left(\frac{\beta\Pi}{2}\right),
\end{eqnarray}
with
\begin{eqnarray}
\label{def-2}
 x_{+}=\sqrt{\frac{(\lambda_{+}-b)}{\Pi}}~~~~~~\mbox{and}~~~~~~
 x_{-}= -\sqrt{\frac{(\lambda_{+}-a)}{\Pi}},
\end{eqnarray}
\begin{eqnarray}
\label{def-3}
&b=&\frac{2\left[J(2\eta+1)-2(D+\gamma_2^{\bot})\right]}{3},\nonumber\\
&a=&\frac{J(-4\eta+1)-2(D+\gamma_2^{\bot})}{3},\nonumber\\
&\Pi=&\sqrt{(J-2(D+\gamma_2^{\bot}))^2+8(J\eta)^2},\nonumber\\
&\kappa=&J-2(D+\gamma_2^{\bot}),\nonumber\\
&\lambda_{\pm}=& \frac{\kappa\pm\Pi}{2}.
\end{eqnarray}
From Eqs. (\ref{Tchpp1}) and (\ref{Tchpp2}) one has the critical temperature for the perpendicular ordering $m_\bot$. 

In addition, the first-order transition lines between the ordered phases (where $m_\parallel\ne0$ and $m_\bot\ne0$) are given when 
the corresponding free-energies are equal, while from the ordered phases and the disordered phase when the free energies 
are the same as the free energy of the paramagnetic phase with $m_\parallel=m_\bot=0$.

\subsection{Analytical Results}
\label{analy}
Although, in this approach, general results can only be achieved through a numerical analysis of the above equations, 
some additional analytical results are available in the limiting case 
$D\rightarrow\infty$. In that limit, we get for the reduced critical temperature ${t_c}=\frac{k_BT_c}{J}$
\begin{equation}
 \label{Tc-analy}
t_c(D\rightarrow\infty)=\frac{2}{\ln\left(\frac{c}{c-2}\right)}.
\end{equation}
Observe that the last expression does not depend on the anisotropy $\eta$, depending only on the coordination number 
$c$ of the hypercube lattice. Therefore, in such limit, quantum effects
are not relevant for the critical behavior. This fact is understandable because when we let $D\rightarrow\infty$ 
in the Hamiltonian (\ref{ham}), the eigenvalues of 
$S^z$ operator can take only the values $1$ and $-1$, since the high energetic cost prohibits that the eigenstates 
associated with zero eigenvalue could be
accessed. Then, the Hamiltonian (\ref{ham}) reduces to the spin-1/2 Ising model, which is a classical one. Equation (\ref{Tc-analy}) gives $t_c=2.885$ in the two-dimensional limit, which should 
be compared to the exact result $t_c=2.269$ \cite{onsager}. For the three-dimensional model one has $t_c=4.932$, 
comparable to Monte Carlo simulations $t_c=4.512$ \cite{PRBLandau}.

In the one-dimensional limit, the present approximation reproduces the exact result for the critical temperature, 
$t_c=0$, even for the anisotropic Heisenberg model in the presence of the crystal field. In addition, for $\eta=0$ 
one further obtains the exact mean value of the quadrupole moment $q$ in one dimension. 
This assure us that, at least for the one-dimensional model, the present approach reproduces the exact results for 
all values of $D$ and $\eta$. One should also say that for the spin-$1/2$ two-dimensional isotropic Heisenberg model 
(where the crystal field is unimportant since it is just a constant in the Hamiltonian) one also gets the exact result 
coming from the Mermin and Wagner theorem $t_c=0$\cite{Mermin}. Despite the fact that for spin-$1$ one does not
reproduce the Mermin and Wagner result for the two-dimensional model, we believe that the comparison depicted
in Table I shows that the present pair approximation is clearly an improvement over the usual mean-field procedure
previously done on this model.
%
%%%%%%%%%%%%%%%%%%%%%%%%%%%%%%%%%%%%%%%%%%%%%%%%%%%%%%%%%%%%%%%%%
\begin{table}
\caption{\it Reduced critical temperatures $t_c$ for the present
model in some limiting cases, for the square ($c=4$) and simple cubic ($c=6$) lattices, according to exact results\cite{onsager,Mermin}, Monte Carlo simulations\cite{PRBLandau},
series expansion\cite{joan,butera,stanley}, the present values and the usual (one-spin) mean-field approach (mostly from reference\cite{khaje}). 
The errors from Monte Carlo and series are in the next (not shown) two digits.
}
\label{tab}
\vskip0.1in
\begin{tabular}{|c|c|c|c|} \hline
  & $c=4$/$c=6$ & present& usual MFA\\ \hline
 \hline
$D\to\infty$ &$$2.269$\cite{onsager}/4.512$\cite{PRBLandau} 
           & $2.885/4.932$ &4/6 \\ \hline
$D=0$, $\eta=0$  &  1.693\cite{joan}/$3.196$\cite{butera}
           &$2.065/3.439$  &  2.667/4    \\ \hline
$D=0$, $\eta=1$     &  0\cite{Mermin}/ 3.000\cite{stanley}
      &  $1.492/2.949$  & 2.667/4  \\ \hline
\end{tabular}
\end{table}
%%%%%%%%%%%%%%%%%%%%%%%%%%%%%%%%
%
%
\subsection{Location of Tricritical and Tetracritical Points}
\label{TCP}
As will be discussed in the next section, the model defined by the Hamiltonian (\ref{ham}) exhibits tricritical and 
tetracritical points in some particular ranges of the parameters of the Hamiltonian. In order to
locate these multicritical points, besides the first- and second-order lines, we have resorted to a Landau like 
expansion of the free-energy (\ref{fe}). For the present case we arrive at the following expansion
\begin{eqnarray}
\label{le}
{\beta}f= f_{0}&+&\frac{1}{2}a_{2}^{\parallel}(t,D,\eta)m_{\parallel}^{2}+
\frac{1}{4}a_{4}^{\parallel}(t,D,\eta)m_{\parallel}^{4}+
\frac{1}{6}a_{6}^{\parallel}m_{\parallel}^{6}\nonumber\\
&+&\frac{1}{2}a_{2}^{\bot}(t,D,\eta)m_{\bot}^{2}+\frac{1}{4}a_{4}^{\bot}(t,D,\eta)m_{\bot}^{4}+
\frac{1}{6}a_{6}^{\bot}m_{\bot}^{6},
\label{le}
\end{eqnarray}
where $f_0$ is a regular function and $a_{2}^{\parallel}$, $a_{4}^{\parallel}$, $a_{6}^{\parallel}$, $a_{2}^{\bot}$, $a_{4}^{\bot}$ 
and $a_{6}^{\bot}$ are coefficients depending on $t,~D, ~\eta$. Tricritical points on the transition lines
separating the parallel ordered and paramagnetic phases  are given by
\begin{eqnarray}                                                                                                      
 a_{2}^{\parallel}(t,D,\eta)=0,~~~~~~~a_{4}^{\parallel}(t,D,\eta)=0~~~~~\mbox{and}~~~~~~a_{6}^{\parallel}>0.\nonumber 
\end{eqnarray}                                                                                                        
The above coefficients have been analytically calculated and their expressions are given in the Appendix.  
On the other hand, tricritical points on the transition lines
separating the perpendicular ordered and paramagnetic phases  are given by
\begin{eqnarray}                                                                                                      
 a_{2}^{\bot}(t,D,\eta)=0,~~~~~~~a_{4}^{\bot}(t,D,\eta)=0~~~~~\mbox{and}~~~~~~a_{6}^{\parallel}>0.\nonumber 
\end{eqnarray}                                                                                                        
It turns out, as we shall see in the next section, that we do not find any tricritical point along this 
transition line.

Finally, the tetracritical point is given when
\begin{eqnarray}
 a_{2}^{\parallel}(t,D,\eta)=a_{2}^{\bot}(t,D,\eta)=0~~~~~\mbox{and}~~~~~~a_{4}^{\parallel}>0, ~ a_{4}^{\bot}>0.\nonumber
\end{eqnarray}                                                                                         
The expression for $a_{2}^{\bot}$ is also given in the Appendix. In this particular
case, the tetracritical point looks like a bicritical point in the temperature 
versus crystal field plane, because we are considering
zero external magnetic fields.

\section{ Numerical Results}
\label{res}
The numerical results of the one-dimensional and the three-dimensional versions of the model
will be presented, including the thermodynamics and the global phase diagrams  as a function
of the parameters of the Hamiltonian. The results for the two-dimensional
model are qualitatively similar to those for the cubic lattice.

\subsection{One-dimensional model}

\begin{figure}[ht]
\includegraphics[angle=0,width=13.5cm]{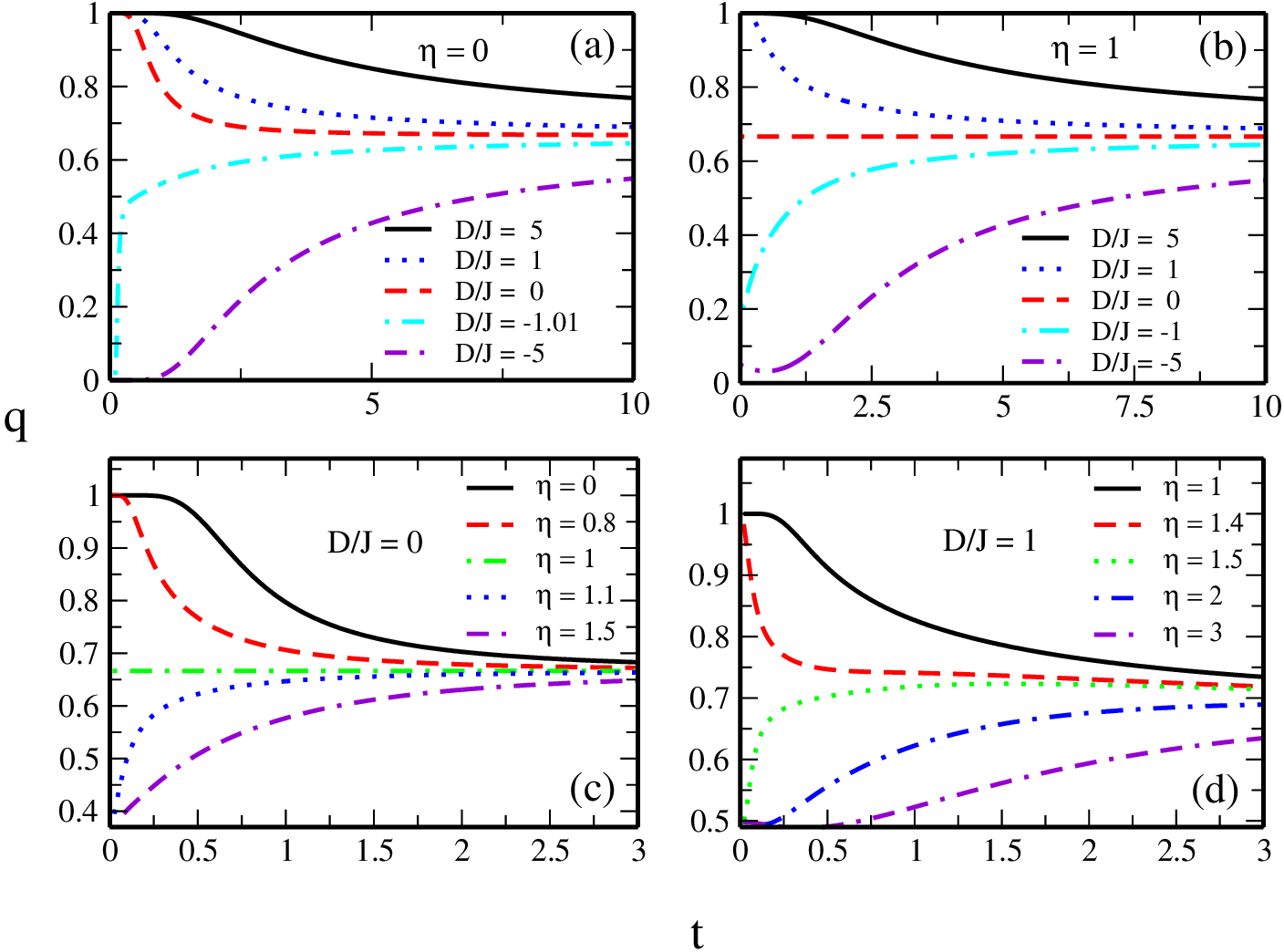} %graph 1%
\caption{\label{q1d} (color online) Quadrupole moment of the one-dimensional model, $c=2$, for several
values of the Hamiltonian parameters.}
\end{figure}

For the one-dimensional model, $c=2$, the above equations give no ordering either along the $z$
axis or in the $xy$ plane, for any value of $\eta$ and $D$. One always has $m_\parallel=m_\bot=0$
with no transition at finite temperatures. This is indeed what one expects for this model. One should say that this is not accomplished by the simple mean-field approach, because it does not distinguish the dimension of the lattice, giving always a finite transition temperature for any value of $c$. Even more recent results obtained from the Green's function method were not able to describe such behavior, and the critical temperature of the one-dimensional model
only vanishes when $\eta=1$ and $D=0$ \cite{song}. Note that when $\eta=0$ one has the classical 
one-dimensional Blume-Capel model which has no phase transition. As we increase $\eta$ from zero, this increasing tends
to destroy the $z$ axis order, which is already disordered, so no transition can be achieved in this case. In addition, for
$\eta=0$ we get the exact free energy and the exact quadrupole moment $q=\langle{S_i^z}^2\rangle$
as obtained from the transfer matrix formalism. The exact quadrupole as a function 
of temperature for $\eta=0$ is shown in Figure \ref{q1d}(a)
for several values of the crystal field. From what has been discussed above, we believe the present results for $\eta>0$, shown in Figures
\ref{q1d}(b)-(d), can be considered near to the exact ones. Unfortunately, in this case, the one-dimensional 
model is non-integrable for spin-$1$ and, up to our knowledge, the results in Figures \ref{q1d}(b)-(d) are novel for the model.  

It is interesting now to analyze the behavior of the one-dimensional quadrupole and see what can be learned from the improved pair approximation.
As Figure \ref{q1d}(b) shows, for the isotropic model, $\eta=1$, the quadrupole is ordered
at zero temperature as soon as one has an easy axis asymmetry for $D/J>0$, while the quadrupole
decreases when one has an easy plane for $D/J<0$. For $D/J=0$, the full isotropic case, the quadrupole 
is always disordered $q=2/3$. The corresponding behavior of $q$ for several values of the anisotropy $\eta$
is shown in Figure \ref{q1d}(c) for $D/J=0$. Here the situation is quite similar to that of
Figure \ref{q1d}{b}, with $\eta<1$ favoring the $z$ axis and $\eta>1$ favoring the $xy$ plane. In
Figure \ref{q1d}(d) for $D/J=1$, which already favors the $z$ axis, one can see a higher value of
$\eta$ (in this case $1.4<\eta<1.5$) in order to favor the $xy$ plane. 
%
%%%%%%%%%%%%%%%%%%%%%%%% fig.1 %%%%%%%%%%%%%%%%%%%%%%%%%%%%%%%%%%%%%%%%%%%%%%% %modificado%
%%%%%%%%%%%%%%%%%%%%%%%%%%%%%%%%%%%%%%%%%%%%%%%%%%%%%%%%%%%%%%%%%%%%%%%%%%%%%% %modificado%
%
\subsection{Three-dimensional model}

In this subsection, we present the numerical results of the behavior of the magnetization $m=\langle{S_{i}^{z}}\rangle$, 
the pair correlation function on the $xy$ plane $\langle{S}_{i}^{x}S_{j}^{x}+S_{j}^{x}S_{j}^{y}\rangle$ and the
global phase diagrams as function of the parameters of the Hamiltonian
for the three-dimensional model. The results for the two-dimensional model are
qualitatively the same. 

\subsubsection{Magnetization and pair correlation function}

\begin{figure}[ht]
\includegraphics[angle=0,width=8.5cm]{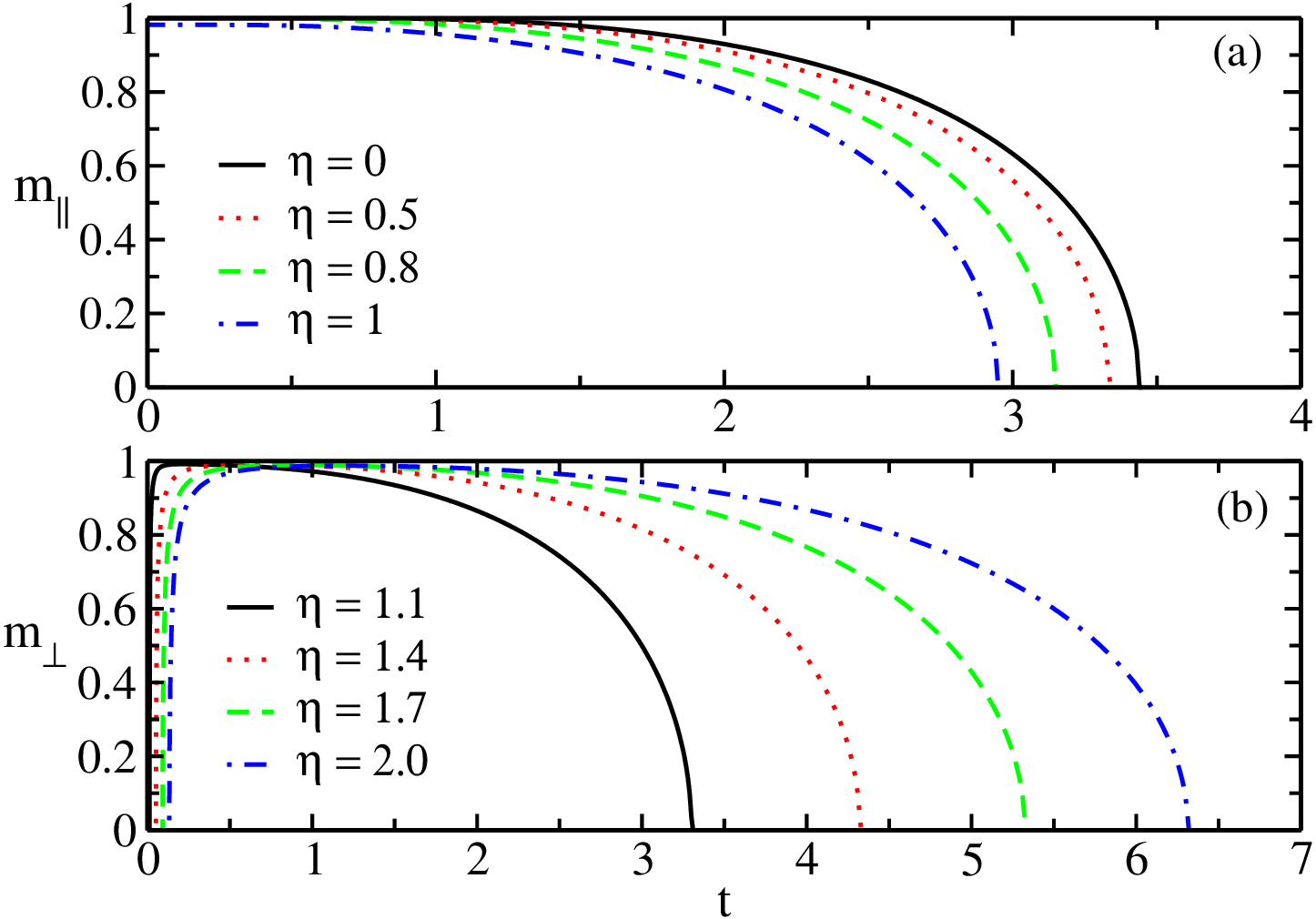} %graph 1%
\caption{\label{mageta} (color online) (a) Parallel $m_\parallel$ and (b) 
perpendicular $m_\bot$ magnetizations as a function of the reduced temperature 
$t={k_{B}T}/{J}$, for $D/J=0$, $c=6$ and several values of the anisotropy $\eta$.}
\end{figure}

\begin{figure}[ht]
\includegraphics[angle=0,width=8.5cm]{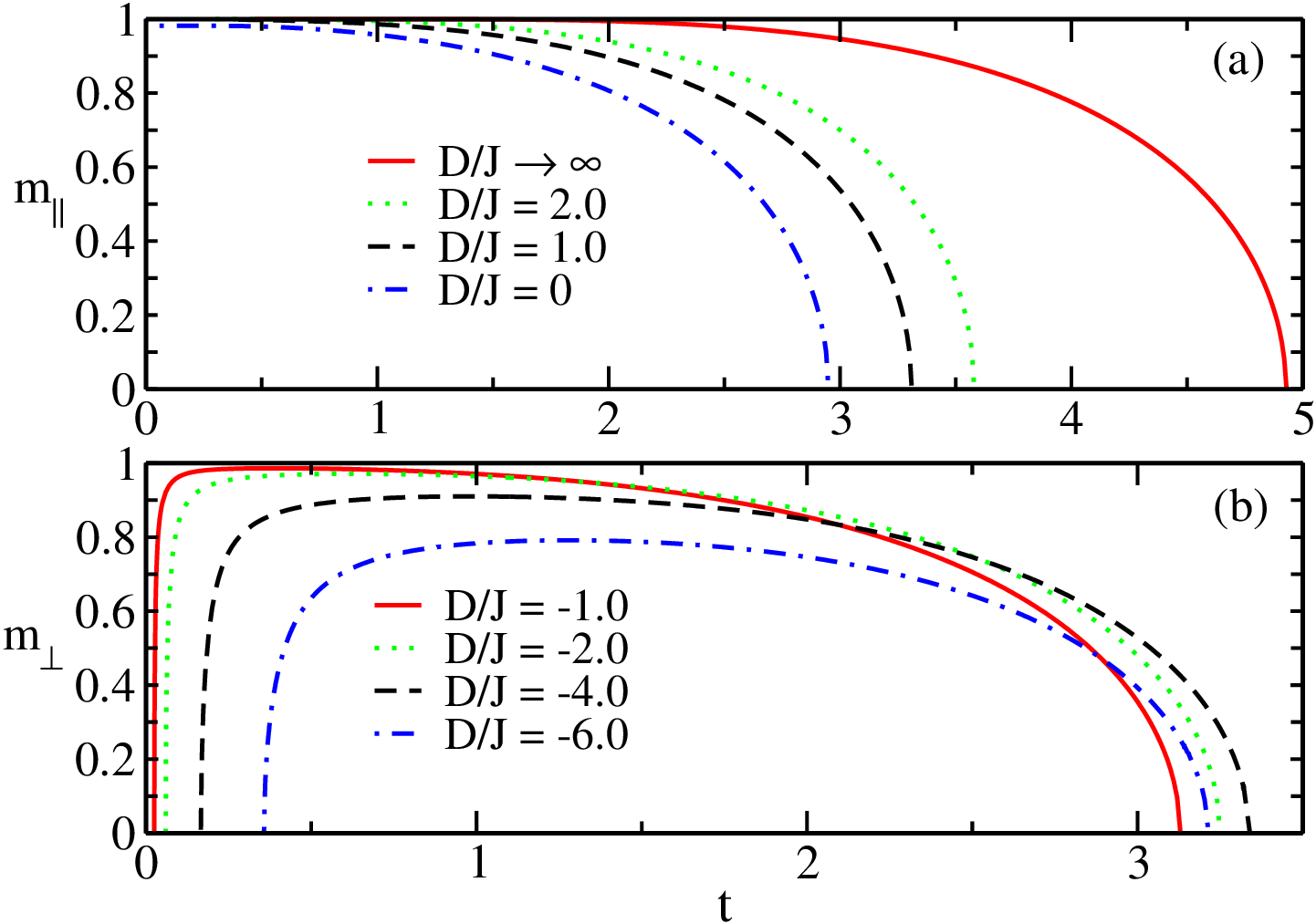} %graph 1%
\caption{\label{magd} (color online) (a) Parallel $m_\parallel$ and (b) perpendicular $m_\bot$ 
magnetizations as a function of the reduced temperature 
$t={k_{B}T}/{J}$, for $\eta=1$, $c=6$ and several values of the crystal field $D/J$.}
\end{figure}

In Figure \ref{mageta} we show the parallel and perpendicular magnetizations as a function
of the reduced temperature $t=k_BT/J$, for several values of $\eta$, for the three-dimensional model and $D/J=0$. 
In $(a)$, we have $\eta<1$ and the stable phase is the one with an Ising like ordering along the $z$ direction
and exhibiting a continuous phase transition as the temperature is increased. One also notes that as the
anisotropy is decreased, the quantum fluctuations increase and the critical temperature is lowered. The spin components tend to lie more
in the $xy$ plane as $\eta\rightarrow1$. On the other hand, in (b), where $\eta>1$, the 
stable phase is the one with a perpendicular ordering. Now, by increasing $\eta$, the easy plane tendency
of the ordering is enhanced and, as a consequence, the critical temperature is also increased. However,
one can see a reentrant behavior where a second continuous transition takes place at low temperatures.
This reentrancy will become clearer when discussing the phase diagrams. Figure \ref{magd} depicts the
magnetizations for $\eta=1$ and various values of the crystal field. In this case, two continuous transitions
are seen for negative values of $D$. These reentrancies are not found in the usual mean-field approach.
%
%%%%%%%%%%%%%%%%%%%%%%%% fig.2 %%%%%%%%%%%%%%%%%%%%%%%%%%%%%%%%%%%%%%%%%%%%%%%
%%%%%%%%%%%%%%%%%%%%%%%%%%%%%%%%%%%%%%%%%%%%%%%%%%%%%%%%%%%%%%%%%%%%%%%%%%%%%%
%
%
%%%%%%%%%%%%%%%%%%%%%%%% fig.3 %%%%%%%%%%%%%%%%%%%%%%%%%%%%%%%%%%%%%%%%%%%%%%%
%%%%%%%%%%%%%%%%%%%%%%%%%%%%%%%%%%%%%%%%%%%%%%%%%%%%%%%%%%%%%%%%%%%%%%%%%%%%%% 
%

\begin{figure}[ht]
\includegraphics[angle=0,width=8.5cm]{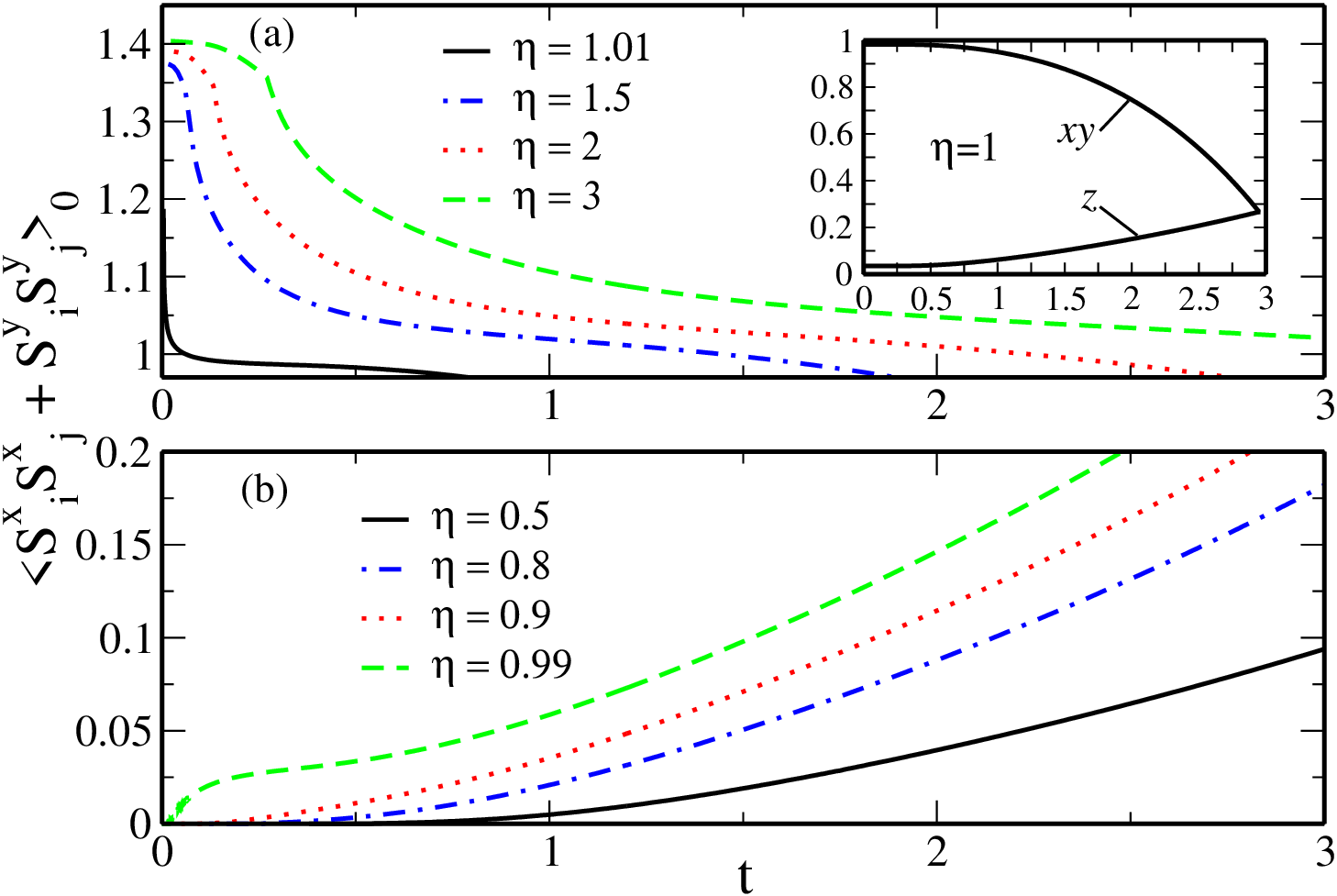} 
\caption{\label{corr} (color online) In-plane nearest-neighbor pair correlation
function as a function of the reduced temperature for the three-dimensional model 
and $D/J=0$ for several values of $\eta$. (a) $\eta>1$ and (b)  $\eta<1$. The inset 
in (a) shows the results for $\eta=1$ for the ordered phase in the $z$ direction
and $xy$ plane, respectively.}
\end{figure}

The nearest-neighbor pair correlation function in the $xy$ plane is shown in
Figure \ref{corr} for $D/J=0$ and various values of $\eta$. For $\eta>1$, the
easy plane situation, this correlation function decreases as the temperature
increases, because the system is already ordered in the plane. On the other
hand, for $\eta<1$, the easy axis case, the in-plane correlation function
increases as the temperature increases, since the temperature tends to destroy
the order along the $z$ direction, favoring in this case the $xy$ plane.  The
inset in Figure \ref{corr}(a) shows the special case $\eta=1$, where we have a
coexistence of both ordered phases, along the $z$ direction and in the $xy$
plane. The pair correlation functions behave differently in each phase, becoming
equal at the tetracritical temperature (see discussion below). 
%
%%%%%%%%%%%%%%%%%%%%%%%% fig.4 %%%%%%%%%%%%%%%%%%%%%%%%%%%%%%%%%%%%%%%%%%%%%%%
%%%%%%%%%%%%%%%%%%%%%%%%%%%%%%%%%%%%%%%%%%%%%%%%%%%%%%%%%%%%%%%%%%%%%%%%%%%%%% 
%

\subsubsection{ Global Phase Diagrams}

\begin{figure}[ht] %modificado%
\includegraphics[clip,angle=0,width=8.6cm]{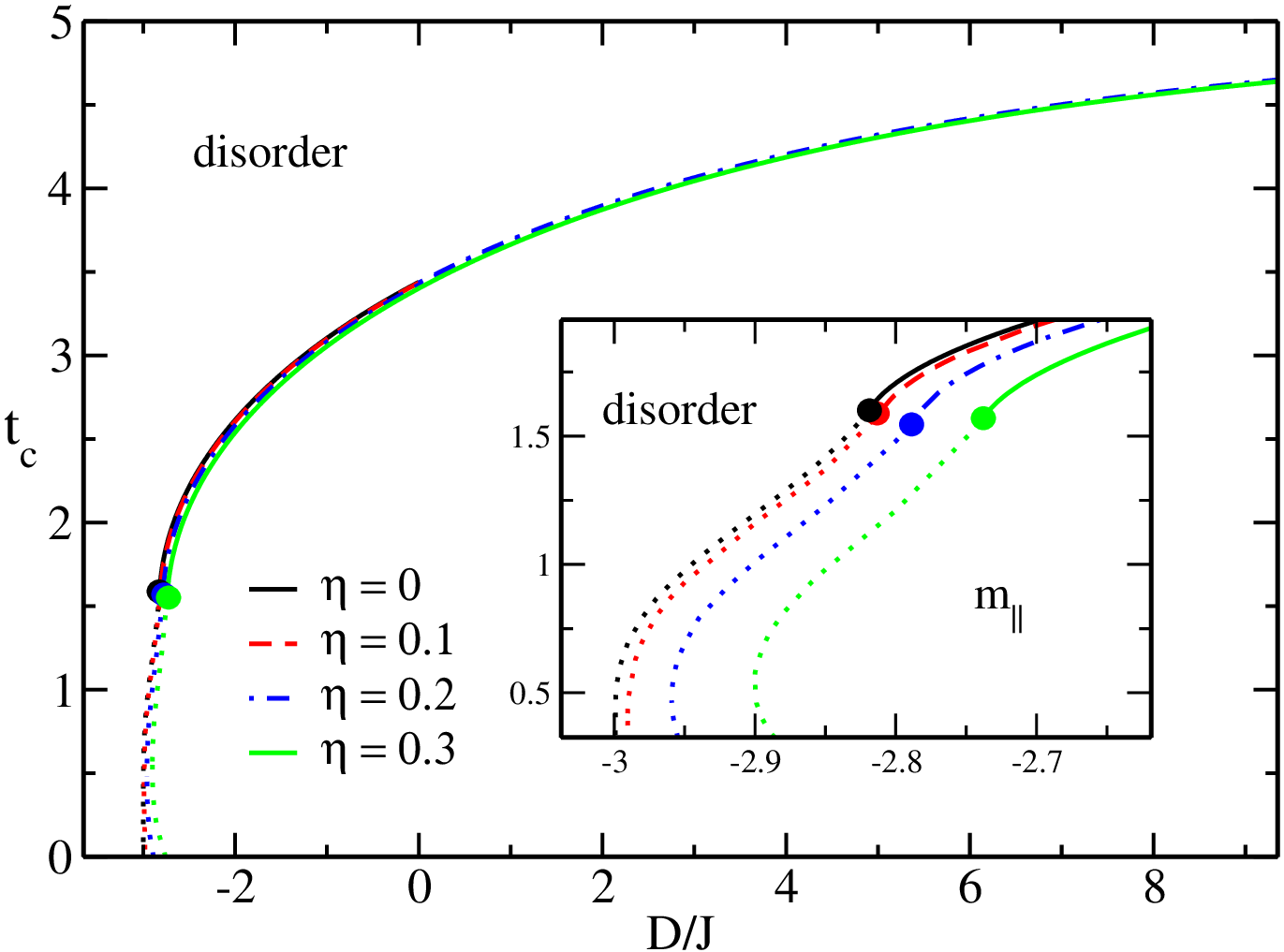} %graph4%
\caption{\label{eta0-0.3} (color online)
Global phase diagram in the reduced temperature versus crystal field for several
values of the anisotropy $\eta\le0.3$, in the three-dimensional lattice with $c=6$. 
The dotted lines refer to
first-order transitions and the others lines (continuous, dashed and dotted-dashed) refer to second-order phase transitions . The circles represent the tricritical points.  The transition is  always
from the stable parallel ordered phase with $m_\parallel\ne0$ to the disordered phase. The inset
shows the low temperature region on a finer scale.
}
\end{figure}

Figure \ref{eta0-0.3} displays the global phase diagram in the reduced critical temperature versus reduced crystal field plane, 
in the three-dimensional limit, for several values of the anisotropy $\eta$. One can see that as soon as 
$\eta<0.33$ the phase
diagram is quite similar to that of the classical Blume-Capel model, presenting second- and first-order transition lines
separated by a tricritical point. In the limit $D/J\rightarrow\infty$ all curves go to the same result 
$t_c=4.932$, as discussed in the text. Apart from the reentrancy at low temperatures, which is clearly depicted in the inset,
for this range of anisotropy the quantum
effects seem not to be enough  to change the character of the transition, and the perpendicular ordered phase is never 
stable. The anisotropy can only stabilize the perpendicular phase when $\eta>0.33$. This is in contrast to the simple mean-field approach, where the perpendicular order is always stable as soon as $\eta>0$\cite{khaje}.
%
%%%%%%%%%%%%%%%%%%%%%%%% fig.5 %%%%%%%%%%%%%%%%%%%%%%%%%%%%%%%%%%%%%%%%%%%%%%%
%%%%%%%%%%%%%%%%%%%%%%%%%%%%%%%%%%%%%%%%%%%%%%%%%%%%%%%%%%%%%%%%%%%%%%%%%%%%%%
%

\begin{figure}[ht] %modificado%
\includegraphics[clip,angle=0,width=8.6cm]{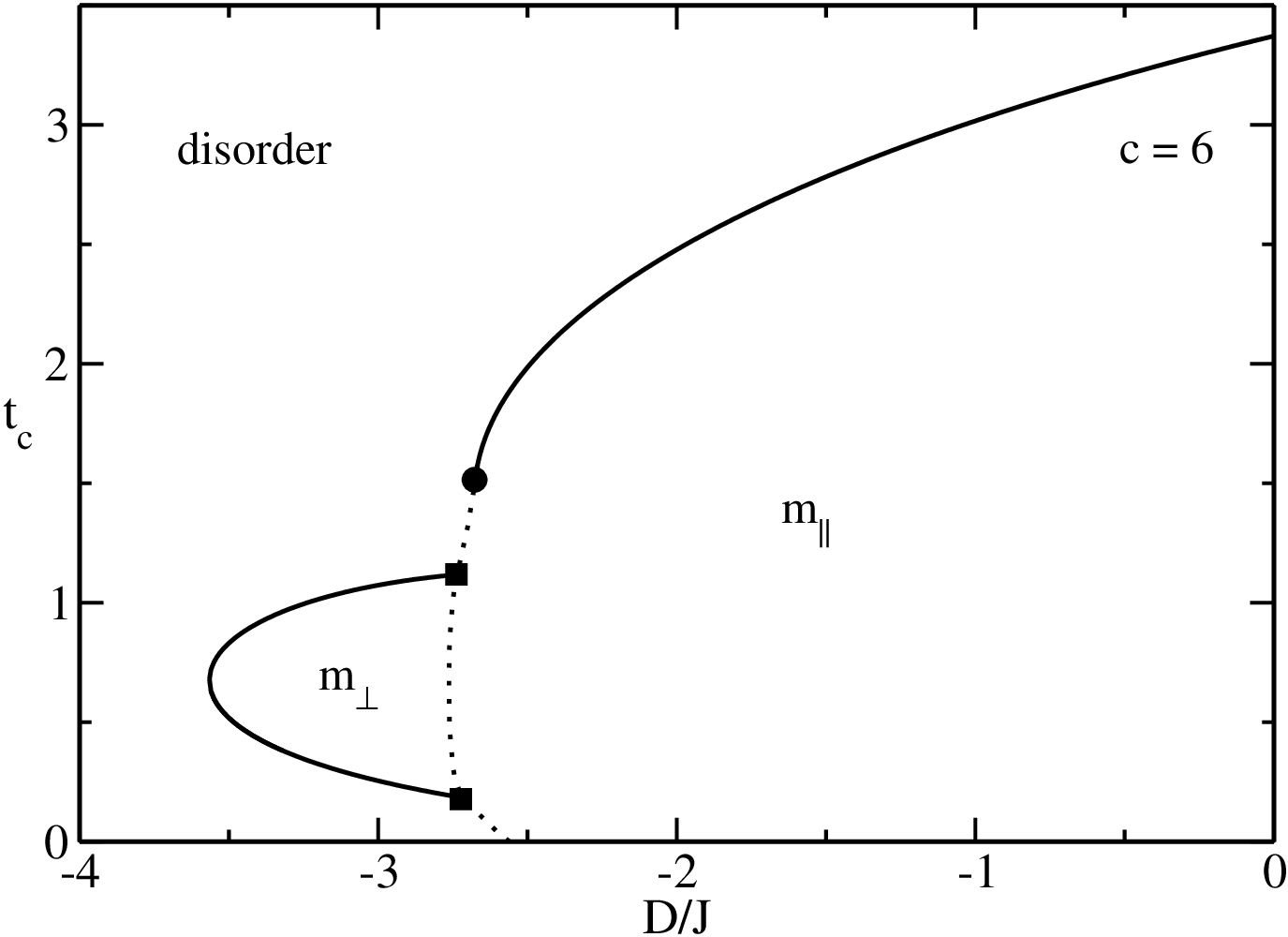} %graph4%
\caption{\label{eta04}
Global phase diagram in the reduced temperature versus crystal field plane for 
 $\eta=0.4$, in the three-dimensional lattice with $c=6$. The continuous
lines refer to second-order phase transitions and the dotted line refers to
first-order transition. The circle represents the tricritical point and the squares
represent the critical endpoints. $m_\parallel$ is the parallel ordered phase and
$m_\bot$ is the perpendicular ordered phase.
}
\end{figure}

\begin{figure}[ht]
\includegraphics[clip,angle=0,width=8.5cm]{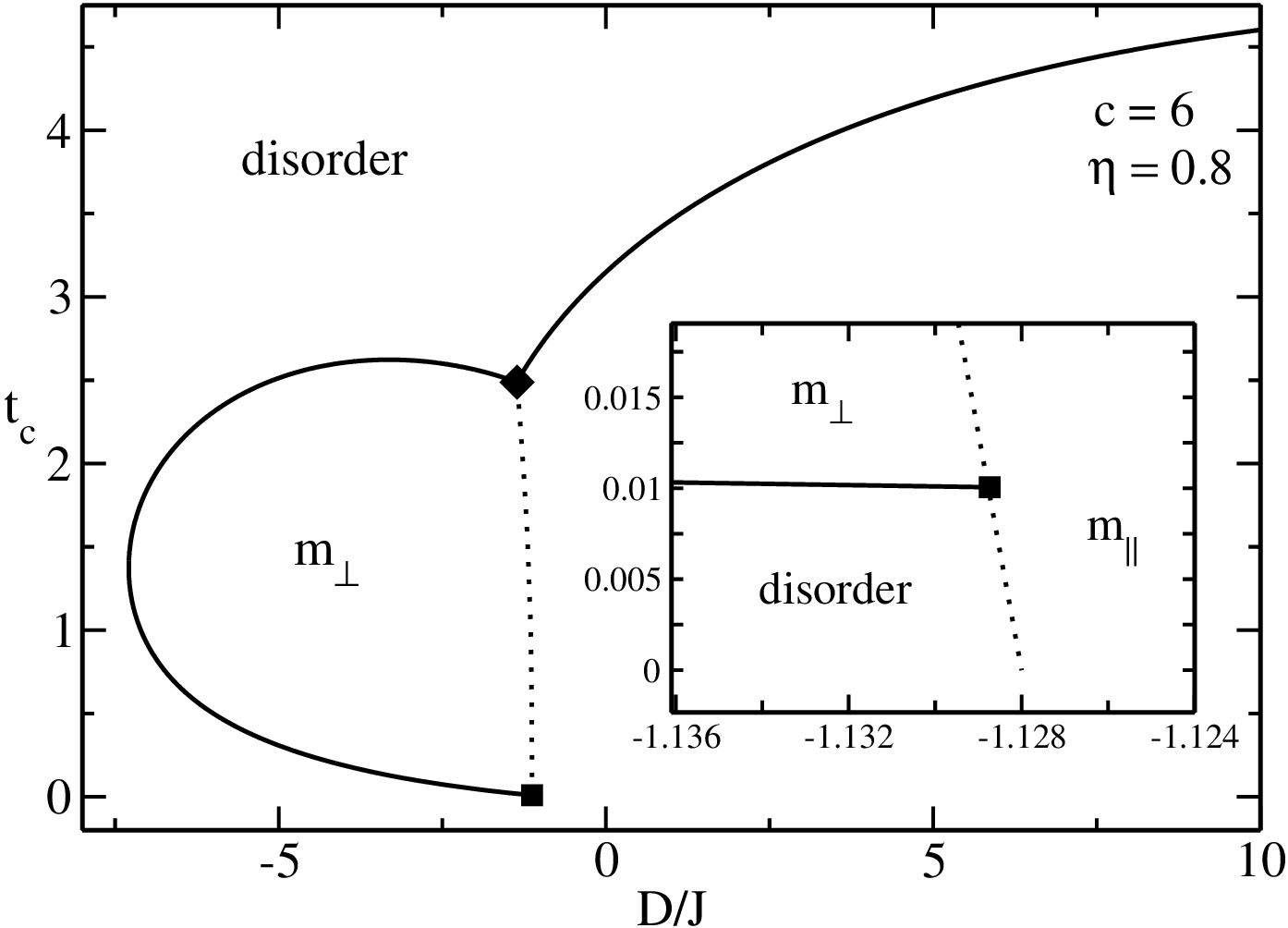} %graph4%
\caption{\label{eta08}
Global phase diagram in the reduced temperature versus crystal field plane for 
 $\eta=0.8$, in the three-dimensional lattice with $c=6$. The continuous
lines refer to second-order phase transitions and the dotted line refers to
first-order transition. The squares represent the low temperature critical endpoint 
and the diamond represents the tetracritical point. $m_\parallel$ is the parallel ordered phase and
$m_\bot$ is the perpendicular ordered phase. The inset shows a closer view of
the low temperature region.}
\end{figure}

For $0.33<\eta<0.49$ the  phase diagram looks 
like  the one shown in Figure \ref{eta04}. In addition to the tricritical point one has two critical endpoints
in the first-order transition line separating the parallel and perpendicular ordered phases. 
%
%%%%%%%%%%%%%%%%%%%%%%%% fig.6 %%%%%%%%%%%%%%%%%%%%%%%%%%%%%%%%%%%%%%%%%%%%%%%
%%%%%%%%%%%%%%%%%%%%%%%%%%%%%%%%%%%%%%%%%%%%%%%%%%%%%%%%%%%%%%%%%%%%%%%%%%%%%%
As $\eta$ increases, the tricritical and the high temperature critical endpoint approaches one another and 
eventually, for $\eta>0.49$, they coalesce in a tetracritical point. 
The phase diagram in this range of anisotropy is depicted in Figure \ref{eta08}.
%
%%%%%%%%%%%%%%%%%%%%%%%% fig.7 %%%%%%%%%%%%%%%%%%%%%%%%%%%%%%%%%%%%%%%%%%%%%%%
%%%%%%%%%%%%%%%%%%%%%%%%%%%%%%%%%%%%%%%%%%%%%%%%%%%%%%%%%%%%%%%%%%%%%%%%%%%%%%
%

\begin{figure}[ht] 
\includegraphics[clip,angle=0,width=8.5cm]{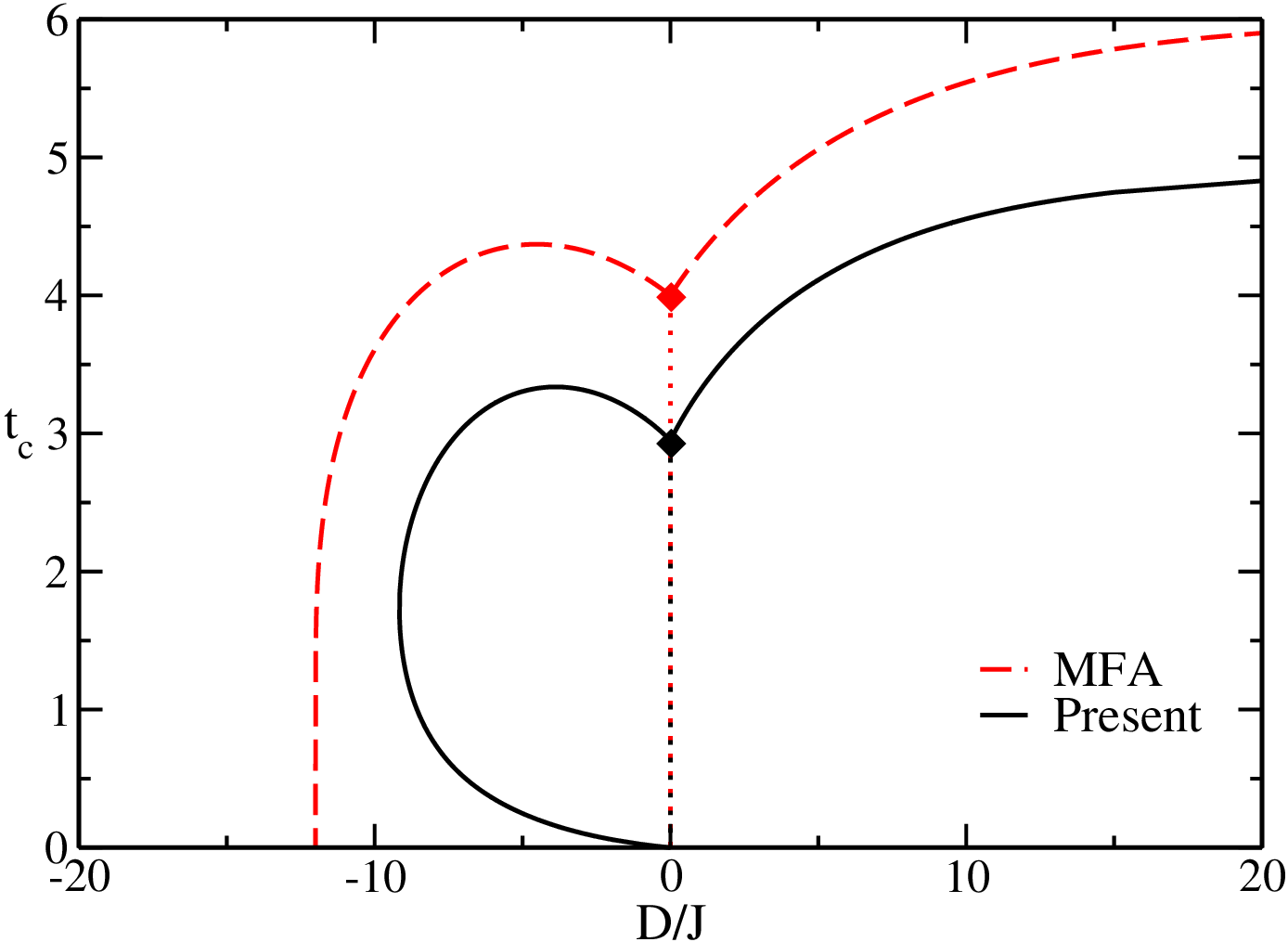} %graph4%
\caption{\label{eta1} (color online)
Phase diagram in the reduced temperature versus crystal field plane for 
 $\eta=1$, in the three-dimensional lattice with $c=6$, according to 
the present approach in comparison to the MFA. The continuous
and dashed lines refer to second-order phase transitions and the dotted lines refer to
first-order transition. The diamonds represent the the tetracritical point.}
\end{figure}

For $\eta=1$ the first-order transition line separating the perpendicular and the parallel phases
is a straight vertical line at $D/J=0$, as shown in Figure \ref{eta1} together with a comparison to the usual
 mean-field approximation (or one-spin approach)\cite{khaje}. Comparing to the mean-field results one can see
that the critical temperature from the pair approximation is systematically below those by taking
just one-spin cluster. The reentrancy only occurs for the pair approximation. Similar results are 
obtained for other values of $\eta>1$, with the slope of the first-order transition line between
the ordered phases becoming positive in this range. 
%
%%%%%%%%%%%%%%%%%%%%%%%% fig.8 %%%%%%%%%%%%%%%%%%%%%%%%%%%%%%%%%%%%%%%%%%%%%%%
%%%%%%%%%%%%%%%%%%%%%%%%%%%%%%%%%%%%%%%%%%%%%%%%%%%%%%%%%%%%%%%%%%%%%%%%%%%%%%

%%%%%%%%%%%%%%%%%%%%%%%% fig.9 %%%%%%%%%%%%%%%%%%%%%%%%%%%%%%%%%%%%%%%%%%%%%%%
\begin{figure}[ht] 
\includegraphics[clip,angle=0,width=8.5cm]{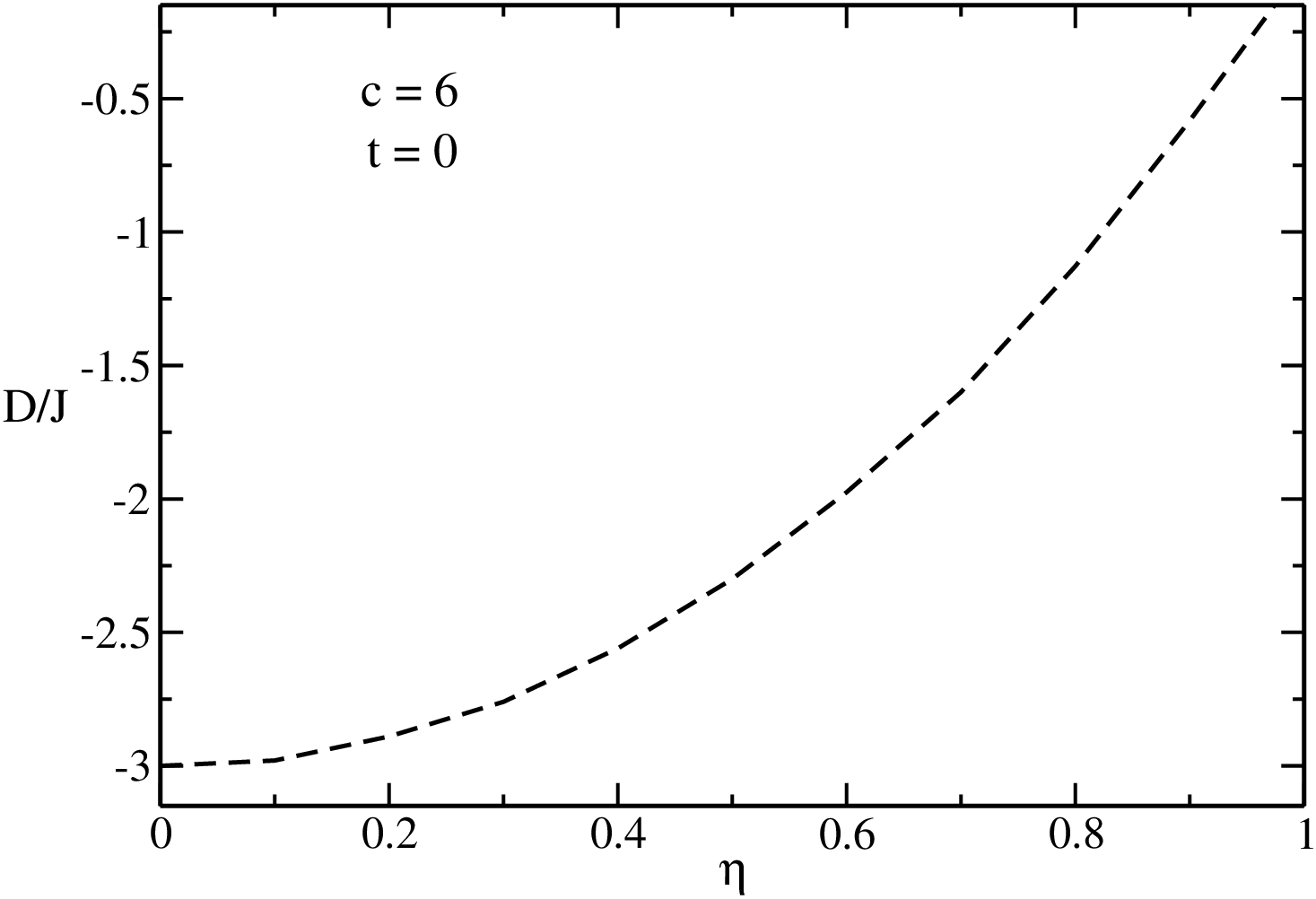} %graph4%
\caption{\label{dxeta}
Reduced crystal field $D/J$ as a function of the anisotropy $\eta$, at zero temperature $T=0$, for the three-dimensional model. }
\end{figure}
%%%%%%%%%%%%%%%%%%%%%%%%%%%%%%%%%%%%%%%%%%%%%%%%%%%%%%%%%%%%%%%%%%%%%%%%%%%%%%
%

All of the above results refer to the three-dimensional model with coordination number $c=6$. The same 
holds for the two-dimensional model. It means that in this case, for spin $S=1$, one does not get the
Mermin and Wagner result $T_c=0$ for $\eta=1$ and $D=0$. However, we believe the approach is suitable to the 
three-dimensional system, mainly when we compare the critical temperatures, as depicted in Table \ref{tab}.

\subsection{Quantum Phase Diagram $T=0$}
\label{PDT}
From Figures \ref{eta0-0.3}-\ref{eta1}, we note that for each anisotropy there exists a value for 
the crystal field $D/J$ in which there is a transition at $t=0$. This transition happens to
be of first order and it is illustrated in Figure \ref{dxeta}. One cannot find neither a 
second-order quantum phase
transition nor a quantum tricritical point  according to the present pair procedure, in contrast to the simple mean-field approach where there is always a quantum phase transition at zero temperature. It should be
stressed, however,  that there is a rigorous proof of the existence of only
first-order phase transitions at low temperature and large anisotropy for the XXZ model \cite{bob}.

\section{concluding remarks}
\label{con}

The anisotropic spin-$1$ XXZ quantum Heisenberg model in a crystal field has been studied according
to a variational approach for the free energy by using a pair approximation. This procedure enables one 
to get the ordering along the $z$ direction and in the $xy$
plane as well. Earlier pair like approaches on the same system could only take into account
the parallel ordering. 

As expected, the one-dimensional model has no phase transition and the
quadrupole moment so obtained are expected to be close to exact one for $\eta>0$. So, contrary to the simple mean-field approximation, the present pair approach can distinguish the dimensionality of the lattice and much more novel information is obtained regarding the free-energy and quadrupole moment for $c=2$.

From the free energies one gets the complete phase diagrams for dimensions greater than one, which are indeed much richer than those from the simple mean-field procedure. The diagrams 
exhibit second-
and first-order transition lines, tricritical and tetracritical points, as well as
critical end points. Although for the spin-$1$ case we do not reproduce the 
Mermin-Wagner theorem in the two-dimensional case, we believe the results are
appropriate for the three-dimensional model. Of course, as it is still a mean-field
approach, it should be necessary more reliable methods to corroborate the 
reentrancies observed in some range of the Hamiltonian parameters. 

\begin{acknowledgments}The authors would like to thank CNPq, FAPEMIG and 
CAPES (Brazilian agencies) for financial support. Fruitful discussions with
S. L. de Queiroz and R. Stinchcombe is also gratefully acknowledged.
\end{acknowledgments}
\appendix
\section{}
In this appendix we present the expressions for the coefficients $a_2^{\parallel}(t,D,\eta)$, $a_{4}^{\parallel}(t,
D,\eta)$ and $a_{2}^{\bot}(t,D,\eta)$ appearing in the Landau                                                                                    
expansion for the free energy (\ref{le}) at the subsection {\ref{TCP}.                                     
Before writing these coefficients, we will define the following variables
\begin{eqnarray}                                                                    
\epsilon&=&\frac{{\cal Z}_{2\parallel}^{p}(0)}{2e^{\beta(D+\gamma_2^{\parallel})}(\cosh(\beta{J}\eta)+ 2e^{\beta(J+D+
\gamma_2^{\parallel})})},\nonumber                                                                                    
\end{eqnarray}                                                                                                        
\begin{eqnarray}                                                                                                      
 A&=&\frac{2e^{\beta(D+\gamma_1^{\parallel})}}{{\cal Z}_{1\parallel}^{f}(0)},\nonumber                               
\end{eqnarray}                                                                                                        
\begin{eqnarray}                                                                                                      
 B&=&\frac{A}{6}-\frac{A^2}{2},\nonumber                                                                              
\end{eqnarray}                                                                                                        
\begin{eqnarray}                                                                                                      
I&=&e^{\beta(D+\gamma_2^{\parallel})}(\cosh(\beta{J}\eta)+8e^{\beta(J+D+\gamma_2^{\parallel})}),\nonumber             
\end{eqnarray}                                                                                                        
where ${\cal Z}_{1\parallel}^{f}(0)$ and ${\cal Z}_{2\parallel}^{p}(0)$ are given by equations (\ref{critic-partition-para-f}) and (\ref{critic-partition-para-p}).

So, we can write
\begin{eqnarray}
 \label{a2}
a_2^{\parallel} &=& \frac{(1-c)}{2A}-\frac{c\epsilon}{2},\nonumber
\end{eqnarray}
\begin{eqnarray}
 a_4^{\parallel}=\frac{B(c-1)}{4A^{4}}-\frac{cI{\epsilon}^{4}}{2{\cal Z}_{2\parallel}^{p}(0)}+\frac{z{\epsilon}^{2}}{
4},\nonumber
\end{eqnarray}
\begin{eqnarray}
\label{critic-equa-h-per}
a_2^{\bot}&=&\frac{2c[e^{\beta(D+\gamma_{1}^{\bot})}-1]}{(1-c)(D+\gamma_1^{\bot}){\cal Z}_{1\bot}^{f}(0)}-
\frac{\sum_{i}e^{-\beta\delta_{i}}\Gamma_{i}}{{\cal Z}_{2\bot}^{p}(0)},\nonumber
\end{eqnarray}
where ${\cal Z}_{1\bot}^{f}(0)$, ${\cal Z}_{2\bot}^{p}(0)$ and $\sum_{i}e^{-\beta\delta_{i}}\Gamma_{i}$ are given by
equations (\ref{def-1}), (\ref{critic-partition-perp-f}), and
(\ref{critic-partition-perp-p}), respectively.

\end{document}